\newcommand{\beq}{\begin{equation}}
\newcommand{\eeq}{\end{equation}}

\newcommand{\gsim}{\lower.7ex\hbox{$\;\stackrel{\textstyle>}{\sim}\;$}}
\newcommand{\lsim}{\lower.7ex\hbox{$\;\stackrel{\textstyle<}{\sim}\;$}}

\documentclass[aps,prl,twocolumn,preprintnumbers,%
               floatfix,nofootinbib]{revtex4}
\usepackage{graphicx}

\def\stacksymbols #1#2#3#4{\def\theguybelow{#2}
    \def\vp{\lower#3pt}
    \def\sp{\baselineskip0pt\lineskip#4pt}
    \mathrel{\mathpalette\intermediary#1}}

\def\intermediary#1#2{\vp\vbox{\sp
     \everycr={}\tabskip0pt
     \halign{$\mathsurround0pt#1\hfil##\hfil$\crcr#2\crcr
              \theguybelow\crcr}}}

\def\to{\rightarrow}

\newcommand{\eqref}[1]{Eq.~(\ref{#1})}

\newcommand{\be}{\begin{equation}}
\newcommand{\ee}{\end{equation}}
\newcommand{\bea}{\begin{eqnarray}}
\newcommand{\eea}{\end{eqnarray}}

\begin{document}


\preprint{MCTP-04-54, hep-th/0409218}

\title{Landscape Cartography: A Coarse Survey of \\
Gauge Group Rank and Stabilization of the Proton}

\author{Jason Kumar and James D. Wells}
\vspace{0.2cm}
\affiliation{
Michigan Center for Theoretical Physics (MCTP) \\
Department of Physics, University of Michigan, Ann Arbor, MI 48109}

\begin{abstract}

The landscape of string/M theory is surveyed over a large
class of type $IIB$ flux compactification vacua.  We derive a simple
formula for the average size of the gauge group rank on the landscape
under assumptions that we clearly state.  We also compute the rank
under the restriction of small cosmological constant, and find a
slight increase.  We discuss how this calculation could impact proton
stability by computing the suppression factor for the number of vacua
with additional gauge group rank that could be used to protect the
proton.  Finally, we present our views on the utility and limitations
of landscape averages, especially in the context of this analysis.

\end{abstract}

\maketitle


\maketitle


\setcounter{equation}{0}

\noindent
{\bf Introduction:}
Recently, there has been renewed interest
in studying mechanisms whereby flux contributions to the superpotential fix
moduli~\cite{flux1,flux2,Dasgupta:1999ss,Giddings:2001yu,Kachru:2003aw}.
There is now a substantial amount of evidence that
such flux vacua are abundant, and there exist
explicit constructions for type $IIB$ compactifications
on orientifolds where all geometric moduli can be
fixed~\cite{Denef:2004dm}.

Although the construction of actual vacua is exceedingly
difficult, it seems that aggregate knowledge of large classes
of vacua are easier to come by.  For example, efforts in this direction
have led to
interesting discussions on the likelihood of low-scale supersymmetry
breaking~\cite{Susskind:2003kw,Robbins:2004hx,Douglas:2004qg,Susskind:2004uv,Dine:2004is,Silverstein:2004sh}.
However, the program should also allow study of issues for which
current data is available, such as the size of the effective theory
gauge group, the number of generations, the stability of the proton, etc.
We primarily focus on this first issue of gauge group rank in this
paper,
and discuss how gauge group rank issues on the landscape
may have relevance to the proton lifetime question.

It is first necessary to consider the type of questions that one
can ask (and hopefully answer) using the string landscape.
One type of question is a ``what" question, such as, what is the
string vacua(um) that describes us?, and
how many vacua do we expect to be close to what we observe in nature?
Another type of question is a ``why" question, namely why we live in this
vacuum and not any other.  This second question often brings
anthropic arguments into play.  We will not rely on such arguments, but
instead emphasize that aggregate analyses on the landscape may help us to sharpen
both questions.

We will focus on a statistical study of the rank of the D3-brane gauge
group on the portion of the landscape described by orientifolded Calabi-Yau
3-fold compactifications of Type $IIB$ string theory.  Our emphasis will
be on quantitative results which are nevertheless applicable to as broad
a class of flux vacua as possible.

In the paragraphs below we will calculate an ensemble average of the gauge group
rank over flux vacua, as a function of the 3-fold parameters.  We will
compute the average rank again in the more restrictive domain of small cosmological
constant.  We will use these averages to
generate estimates for the percentage of vacua that could plausibly
stabilize the proton from extra gauge symmetries.
We will then discuss at more length the interpretation
of these results, and close with some ideas for future
development.

\bigskip
\noindent
{\bf Average Gauge Group Rank:}
We will consider flux vacua of an orientifolded Calabi-Yau 3-fold compactification
of Type $IIB$ string theory.  Our aim will be to compute the ``average" rank of the
D3-brane gauge group for any choice $Y$ of Calabi-Yau 3-fold.

We let $X$ denote a Calabi-Yau 4-fold such that the orientifold limit of
F-theory compactified on $X$ is Type $IIB$ compactified on the orientifold
of $Y$.  The tadpole cancellation condition is
\begin{equation}
L_* ={\chi(X) \over 24} =N_{D3} + \int F^{RR} \wedge H^{NS}
\end{equation}
where $\chi$ is the Euler character and $N_{D3}$ is the net D3-brane charge.
For any particular choice of $Y$ plus orientifold action
(or equivalently, $X$), $L_*$ will be fixed.
But clearly $N_{D3}$ will need to vary with the choice of fluxes for each vacuum.  We
will calculate an ensemble average of $N_{D3}$ over all flux vacua.

We quickly review the conventions and notation we use, which follow
those of~\cite{Ashok:2003gk,Denef:2004ze}.  If $n$ is the number of complex structure
moduli of $Y$ which are not projected out by the orientifold, then
the number of independent fluxes which may be turned on is $2m=2n+2$.
As argued in \cite{Gukov:1999ya}, the superpotential can be written as
\be
W =\int_M G \wedge \Omega (z)
\ee
where $G_3 =F^{RR}-\tau H^{NS}$ and the $z$ are the complex structure
moduli.  The perturbative superpotential is thus determined by the fluxes,
and it is indeed possible to invert this relationship and define a basis
for the fluxes whose coefficients are determined by the superpotential.

The coefficients are written as
\begin{eqnarray}
W &=& X
\nonumber\\
D_A W &=& Y_A
\nonumber\\
D_0 D_I W &=& Z_I
\end{eqnarray}
where $A=1 \ldots n+1$, $I=1 \ldots n$.

The tadpole condition can now be rewritten as
$L_* =N_{D3} + L $ where
\begin{eqnarray}
L &=& \int F^{RR} \wedge H^{NS} = |X|^2 -|Y|^2 +|Z|^2
\end{eqnarray}
Note that although $X$, $Y$ and $Z$ are generically not
quantized, $L$ itself is an integer quantized in string units.  $X$, $Y$ and $Z$
are determined by the choice of fluxes $\overrightarrow{N}$ in the
integral basis and the complex
structure moduli $z$, so that for any choice of $z$ they should be discrete,
though not necessarily quantized.

The number of supersymmetric flux vacua
(by which we mean that supersymmetry is not broken at tree-level by the fluxes,
though it can be broken by non-perturbative dynamics) satisfying
the tadpole condition is~\cite{Denef:2004ze}\footnote{See also
\cite{Giryavets:2004zr}, where power law dependence on $L_*$ is
verified in a particular model.  Note also that $L_*$ must be somewhat
larger than $n$ in order for the integration to give a good approximation
to a discrete sum over fluxes.  This condition will be satisfied for
a wide class of models, including ones which are phenomenologically
interesting.}
\begin{eqnarray}
N (L \leq L_*)&=&
{(2\pi L_*)^{2m} \over (2m)!} |\det \eta|^{-{1\over 2}} \int d^{2m} z
\det g \rho (z) \nonumber \\
\end{eqnarray}
where $\rho $ is the flux vacua density on the complex structure moduli
space, $g$ is the metric on moduli space, and $\eta$ is the Jacobian
for the change of variables from integer fluxes $\overrightarrow{N}$ to
$X$, $Y$ and $Z$.
We will not discuss these factors further (referring
the reader instead to \cite{Denef:2004ze}), since these factors will
not be essential for us.

What we are interested in is the number of vacua for which
$N_{D3}=L_* -L \geq R_0$.  This is given by
\begin{eqnarray}
N (R_0)&=&
{(2\pi )^{2n+2} \over (2n+2)!}
(L_* -R_0)^{2n+2}
|\det \eta|^{-{1\over 2}} \nonumber \\ & & \times \int d^{2m} z
\det g \rho (z)
= c_Y (L_* -R_0 )^{2n+2}
\end{eqnarray}
where $R_0$ is the lower-bound on the gauge group rank, subject
to some caveats that we explain shortly.
We thus find the ``rank density" of flux
\bea
\rho (R)&=& {\partial N(R) \over  \partial R}
\nonumber\\
&=& (2n+2) c_Y (L_* -R )^{2n+1}
\eea

An average rank of the D3-brane gauge group
can then be computed
\bea
\langle R_{D3} \rangle &=& {1\over N(0)}\int_0 ^{L_*} dR \, \rho (R) R
\nonumber\\
&=& {1\over c_Y L_* ^{2n+2}} {c_Y L_* ^{2n+3} \over 2n+3}
\nonumber\\
&=& {L_* \over 2n+3}
\label{eq:RD3}
\eea
Note that all dependence on the geometric data of the Calabi-Yau
is contained in the constant $c_Y$, which factors out.  Thus
the average size of the gauge group depends only on the number
of complex structure moduli and the Euler character of the 4-fold,
but not on the detailed structure of the Calabi-Yau.

One subtlety here is a possible
additional $c^N$ degeneracy factor that would be
expected\footnote{We thank M. Douglas for discussions on this point.}
for a gauge theory arising from $N$ branes~\cite{Douglas:2003um}.
This degeneracy comes
from the multiplicity of vacua describing different stabilizations
of matter, and is thus in the category of ``gauge dynamics''
that ultimately determines the preserved gauge group rank in the
low-energy limit. Our accounting of rank does not take into account
the various subsequent symmetry breaking patterns giving rise to distributed
preserved gauge groups with rank $R_i$ less than $R=N_{D3}$.

In the simplest case imaginable, where $Y$ is
Kthe standard $T^6 /Z_2$ orientifold with symmetric fluxes, one
finds $n=1$ and $L_* =16$.  This yields
$\langle R_{D3}\rangle= {16 \over 5}$,
which of course is close to the SM gauge group rank of 4.
One should not take this specific result literally for phenomenology, and we
mention it only as a curiosity.  It may be suggestive of a more general
result over a large class of manifolds, and certainly gives hope
that searches for quasi-realistic
string models on type $IIB$ orientifold backgrounds with flux
may generically have
gauge group rank near that of the SM.


It is important to interpret the gauge group rank average carefully.
It is \textit{not} a prediction for what the rank of the D3-brane gauge group
must be if the real world is represented by an orientifold of Type $IIB$
on a Calabi-Yau 3-fold.  It is computed in the ensemble where each flux
vacuum is given equal weight, and it is not clear that such an
ensemble is correct for the purposes of vacuum selection.  We will discuss
more thoroughly the uses of this type of average later in this paper.

One should also keep in mind that, strictly speaking, we have not computed
the average rank of the D3-brane gauge
group.  It is more properly the average net D3-brane charge.
We might add several
$D3/\overline{D3}$
pairs, which will change the rank without changing the net charge.
However, these states are at best meta-stable, and will decay to a state
with flux,
but no anti-branes~\cite{Kachru:2002gs}.
In principle, we should consider such states that have a
lifetime of cosmological scales, but for this coarse survey we
can ignore them.

It is interesting to note that the rank of the D7-brane gauge
group is given by \cite{Gopakumar:1996mu}

\bea
R_{D7} &=& h^{1,1}(X) -h^{1,1}(B)-1
\eea
where $B$ is the 3-fold base of the elliptically
fibered Calabi-Yau 4-fold $X$.  Both the average D3-brane
and D7-brane gauge group rank can be written as simple
functions of the Betti numbers of the Calabi-Yau 4-fold and
its base.  Note, however, that the D7-brane gauge group rank
is fixed for a given choice of Calabi-Yau, and is independent
of the choice of flux. Also note its dependence on K\"ahler moduli,
as opposed to complex structure moduli in the case of the
D3-brane rank.

\bigskip
\noindent
{\bf Impact of Small Cosmological Constant:}
We would like to study the topography of the landscape
in the limit of small cosmological constant.
In particular, we would like to consider the average D3-brane
gauge group rank in the small $\lambda $ limit.
Our computation is greatly aided by the underlying analysis
of~\cite{Denef:2004ze}.  They showed that the number
of supersymmetric vacua that satisfy the tadpole condition
and $\lambda =|W|^2 \leq \lambda_*$ is given by
\footnote{This distribution reflects only the contribution to
the vacuum energy from the flux potential and from
the Kahler potential
derived from special geometry.  One expects corrections, but under
the assumption that these corrections are random, the final
distribution should be somewhat similar.  The distribution used
here would then be a good toy model.  }
\begin{eqnarray}
N_{\lambda_*} &=&
{(2\pi L_*)^{2n+2} \over (2n+2)!} |\det \eta|^{-{1\over 2}}
\int_M d^{2n+2} z g \int_0 ^{\lambda_*} d\lambda
\rho(\lambda, z) \nonumber \\
\end{eqnarray}

In the limit where $\lambda_* \ll L_*$ (the small cosmological constant limit),
this can be simplified to
\begin{eqnarray}
\lim_{\lambda_*\ll L_*} N_{\lambda_*} &\to & 2\pi {(2 L_*)^{2n+1} \over (2n+1)!}
\lambda_* |\det \eta|^{-{1\over 2}} 
\int_M d^{2n+2} z g I({\cal F})
\nonumber\\
&  = & b_y L_* ^{2n+1} \lambda_*
\end{eqnarray}
We then find that the number of flux vacua with $N_{D3} >R_0$ and
$\lambda \leq \lambda_* \ll L_* -R_0$ is
\bea
N_{\lambda_*} (R_0 ) &=& b_y (L_* -R_0 )^{2n+1} \lambda_*
\eea

As before, we can compute the rank density of flux vacua and use
this to compute the average rank.  We find

\begin{eqnarray}
\rho_{\lambda_*} (R) &=& (2n+1) (L_* -R )^{2n} \lambda_*
\nonumber\\
\langle R_{D3}\rangle_{\lambda_*} &=&
{1\over N_{\lambda_*}(0)}\int_0 ^{L_*} dR \,
\rho_{\lambda_*} (R) R
\nonumber\\
&=& {L_* \over 2n+2}
\end{eqnarray}
Note again that this ensemble average does not depend on
the geometric data of the Calabi-Yau, but only on the Euler
character of the 4-fold and on the number of complex structure
moduli.
For the $T^6 /Z_2$ orientifold model $\langle R_{D3}\rangle_{\lambda_*} =4$,
which again is curiously close to (same as) the SM rank.

\bigskip
\noindent
{\bf Stability of the Proton:}
The proton lives longer than $10^{32}$ years.  Explaining why this
must be so in a more fundamental theory is a major research challenge.
The survey on gauge group rank performed above has relevance to this
issue.

The renormalizable operators of the SM effective theory
forbid baryon number interactions due to an accidental symmetry
resulting from the restricted particle content.
In string models with
low  or intermediate scales (including string scale), which many type $IIB$ flux
vacua apparently
have~\cite{Camara:2003ku,Marchesano:2004yq,Marchesano:2004xz},
accidental symmetries are not enough to protect
the proton, as a bevy of induced higher-dimensional operators would
generically destabilize the proton.

In supersymmetric effective
theories, proton destabilizing operators are present even at the
renormalizable level.  It is likely that a symmetry would be at play
in these circumstances, since
assuming that every one of the many
dangerous operators had a tiny coefficient by accident
would be hard to fathom.

The most commonly assumed symmetry to stabilize the proton in supersymmetry
is a $Z_2$ $R$-parity.
One often takes $R$-parity for granted in low-scale
model building and implicitly imposes it as a global symmetry
on the theory.  However, treating global symmetries as fundamental
is disfavored in string/M theory.

Despite the allusion to $R$-symmetry, $R$-parity can be interpreted
entirely as a $Z_2$ matter parity on the
chiral superfields. The $Z_2$ in turn can be thought of as arising from
a discrete subgroup of a gauged symmetry,
$G\supset U(1)_{B-L}\supset Z_2$,
that is broken by a condensing scalar carrying the appropriate
charge~\cite{Krauss:1988zc,Martin:1992mq,Goh:2003nv}.
The resulting $Z_2$ symmetry is discrete-gauge anomaly
free~\cite{Ibanez:1991pr} on the MSSM
particle content, as is expected
since $U(1)_{B-L}$ is also anomaly free.
Other discrete symmetries in addition to the $Z_2$ $R$-parity
could also stabilize the proton and come from higher gauge group
symmetry breaking~\cite{Ibanez:1991pr}.

The above discussion suggests that the rank of the gauge group of
nature may need to be at least one step higher than that of the SM
in order to stabilize the proton using gauge symmetries, and possibly
many steps higher. For example, nature may have chosen $SO(10)$ unification
with an embedded $R$-parity.  Given the conventional assignments of the
MSSM states, the lowest dimensional representation of
$SO(10)$ that can condense to give $R$-parity is the
${\bf 126}$~\cite{Martin:1992mq,Goh:2003nv,Aulakh:2000sn}.
Such a high-index field is not easy to
obtain in these stacked $D$-brane models,
since we expect lower-index bifundamentals. (High-index
$SO(10)$ fields
are hard to obtain in other constructions as well~\cite{Dienes:1996du}.)
In this case, the
starting-point gauge group  would need significantly higher rank.
Brane separation could then Higgs this high-rank group down to $SO(10)$,
decomposing the bifundamentals of the group into the {\bf 126} of $SO(10)$
(and other needed states).  All unwanted states would need to be lifted
by normal Higgs mechanisms or projected out of the
effective theory by judicious choices of the compactification.

The suggestion that gauge symmetries are at the origin of proton stabilization
gains even more strength when we consider
the generalized Green-Schwarz mechanism~\cite{Ibanez:1998qp}
of type $IIB$ theories, which
admits additional pseudo-anomalous $U(1)$'s
that would otherwise look unacceptable from an effective
field theory point of view.  This enhanced set of $U(1)$
theories is at
nature's disposal to stabilize the proton~\cite{Ibanez:1999it},
and the hypothesis that
protons are stabilized by an additional gauge group looks quite promising.

The computation of gauge group rank now becomes relevant to proton
decay in this context.
We would therefore like to compute the percentage of susy flux vacua (for
any given choice of CY compactification) that can allow for an
extra $U(1)$ on the D3-branes.  If $R_{sm}$ is the rank of the
SM gauge group, then we can rephrase this by asking
what fraction of vacua
contain D3-branes such that $N_{D3}=R>R_{sm}$.  This is the
same as the number of flux vacua with $L< L_* -R_{sm}$.

As shown earlier, this number is given by
$N=c_y (L_* -R_{sm})^{2n+2}$. From this we see
that the fraction of all susy vacua that
have $R$ larger than the SM group is given by
\begin{equation}
\eta ={(L_* -R_{sm})^{2n+2} \over L_* ^{2n+2}}
=\left( 1-{R_{sm} \over L_*}\right)^{2n+2}
\label{eq:RsmL}
\end{equation}

We will assume that we are in the limit where the number of
unprojected complex structure moduli is large, $n \gg 1$.
Then $\langle R\rangle ={L_* \over 2n+3} \sim {L_* \over 2n+2}$, where
$\langle R\rangle =\langle N_{D3}\rangle$ is the
average rank of the gauge group (possibly, unification gauge group)
for that choice of CY compactification.
We then see that
\begin{equation}
\eta \sim \left(1-{R_{sm} \over \langle R\rangle (2n+2)}\right)^{2n+2}
\sim e^{-{R_{sm}/ \langle R\rangle}}
\end{equation}
in the high $n$ limit.  In practice, as long as
$R_{sm}/\langle R\rangle\sim 1$ we only require that
$2n+2\gg 1$ (e.g., even $n=1$ would suffice) for the above
exponential formula to be a good approximation.

This is the fraction of susy vacua with D3-brane gauge
group rank greater than that of the SM.  Note that
it only depends on $\langle R\rangle$ in the limit of a large number of
moduli.  It does not depend on the details of the CY moduli
space, on any singularities, or even on the Euler character
of the relevant 4-fold.

Earlier we found that $\langle R\rangle =16/5$ (or $4$, when restricting
to small $\lambda_*$) for the $T^6/Z_2$ orientifold example,
which implies that the suppression price one
pays for having a group with rank higher than the SM gauge group
is not more than a factor of 5 in the approximation.
This relatively low suppression factor is
interesting\footnote{If the
minimal rank $R_{M}$ of the unified theory needs to be
significantly above $R_{sm}$,
as might be required by $SO(10)$ with
a ${\bf 126}$ representation,
the compactification manifold would need
a high-enough Euler number to reach the requisite $L_*$.  Eq.~(\ref{eq:RsmL})
would still
be valid except $R_{sm}$ would be replaced by $R_{M}-1$.}. It could
be indicative of a fruitful direction in string model-building: vacua
that stabilize the proton with additional gauge symmetries may
be more stringy/landscape natural than ones that utilize
intrinsic discrete symmetries.

This very tentative supposition emerges partly from taking
into consideration the
analysis of~\cite{Banks:2003es}.  Although
$T^6/Z_2$ orientifold compactifications of type $IIB$ theories
may have large discrete symmetries, the fluxes typically break
them all. Very large exponential suppression
factors result.  It was later suggested that this analysis is based
on models that might not be realistic enough to draw definitive
conclusions~\cite{Dine:2004is}, and discrete symmetries might be more abundant
than originally thought~\cite{Kachru:Aspen}.

Nevertheless, the landscape terrain of~\cite{Banks:2003es}
is very similar to the landscape terrain we are considering here.
On this terrain, extra gauge symmetries that
might have a chance to stabilize the proton are perhaps more copious
than extra stringy discrete symmetries that might have a chance
to stabilize the proton.

\bigskip
\noindent
{\bf Utility of Landscape Averages:}
Perhaps as important as a survey of the landscape is the
question of what to do with this information.  Landscape
averages of the form computed above do \textit{not} necessarily
provide predictions of the real world.  Similarly, a failure
of an average to match experimental data would not falsify the
landscape, let alone string theory.  How landscape data could
be used depends on the question one asks.

For example, if one is attempting to build a specific string
model that matches the real world, then landscape statistics
are useful in determining ``good" criteria.  Suppose one decides
to look for a model that has the experimentally determined
properties $P_1$, ..., $P_n$ as a candidate for the real
world\footnote{These properties are similar to what Dine, Gorbatov
and Thomas~\cite{Dine:2004is}
refer to as priors, though we use them in a different context.}.
If these properties are in fact rather generic on the landscape,
they would not provide good search criteria for selecting
a model to study.  In particular, if given properties
$P_1$, ..., $P_{n-1}$, one finds that property $P_n$ is rather
generic, then $P_n$ provides little useful information and is not
a good search criterion.  Instead, one would hope that a
sophisticated understanding of the landscape would identify
search criteria that are both tractable for a model-builder and
highly non-generic\footnote{To emphasize, genericness on the landscape
is neither ``good'' nor ``bad'' in our view.  It is only useful or not
useful depending on the question one is trying to answer.},
as this would imply that a model that had
these properties is more likely to be ``the right one."

One might instead be interested in understanding what, if any,
principles determine the selection of the vacuum.  In this case,
an understanding of the statistics of the landscape would allow
one to identify which properties actually \textit{require} such
a Vacuum Selection Principle.  For example, suppose it is found
that, given experimentally determined properties
$P_1$, ..., $P_{n-1}$, the property $P_n$ appears generically
on the landscape.  From this, one might conclude that the appearance
of property $P_n$ in the real world does not require an independent
explanation (although it may in fact have one); it is simply a very
generic result on the space of vacua.

One the other hand, it is important to note that the landscape
does not require all experimentally measured features to be
generic.  On the contrary, it seems clear that many
properties will be non-generic given a set of prior properties.
In those cases, one would need to find a principle that selects
the non-generic properties. The anthropic principle could be
viewed as a possible such principle, but one could certainly imagine
that other principles would emerge.

\bigskip
\noindent
{\bf Conclusions:}
In this work, we have made some rather simple calculations of ensemble
averages of quantities to gain insight into the average
rank of the D3-brane gauge group in
flux compactifications.  We found that the average rank can be
computed quite generally, and has very little dependence on the detailed
geometric data of the Calabi-Yau.  We also found that employing
extra gauge symmetry to protect the proton is an apparently reasonable
supposition on the landscape, as every increment in gauge group rank
does not generate a huge suppression factor on the landscape.

Given that we have computed averages of the D3-brane gauge
group rank, one would like to know how this relates to phenomenology.
One natural hope would be that the D3-brane gauge group corresponds
to some unification group.
Note, however, that there are several other mechanisms by which
non-abelian gauge groups can appear.  For example, these gauge groups
can arise from D7-branes (which were discussed earlier) as well
as from the enhancement of gauge symmetry due to D-branes wrapping
shrinking cycles of the Calabi-Yau.
Thus, it could also be that the visible sector unification group
arises from some other mechanism, while the D3-branes generate a hidden
sector gauge group.  Either way, these orientifold constructions will
contain some sector given by the gauge theory on these D3-branes.  Independent
of how we might want to interpret the D3-brane sector in phenomenology,
it seems likely that there is a larger class of vacua which are dual
to these models, and thus contain a gauge group (arising from various
mechanisms) which is dual to the D3-brane group.

This limited success suggests more avenues of study.  It
would be interesting to examine ensemble averages of other
quantities, such as $R$-symmetry breaking parameters and supersymmetry
breaking parameters in regimes where supersymmetry is broken at
tree level by fluxes.  This data could be useful for examining the
possible appearance of interesting effects like intermediate-scale
supersymmetry breaking and proton stabilization by means other than
the pure gauge group discussion we presented here.

It would also be useful to consider large classes of orientifolded
Calabi-Yau 3-fold compactifications
to determine if there are any trends that persist in a more
encompassing averaging procedure.
Another way to view this problem is to
to perform our survey along the lines discussed above, but with an
additional averaging over a large class of Calabi-Yau 4-folds
in F-theory compactifications~\cite{Sen:1997gv}, with well-enumerated
moduli that survive the orientifold projections.  In a sense,
this would require including apppropriately normalized integrals
over the $L_*$ and the $n$ complex structure moduli of
eq.~(\ref{eq:RD3}).

The calculations here are at the very beginning of a potentially
interesting road.   With a much more detailed survey of the landscape,
one might hope to
be able to provide useful input to model-builders, as well as sharpen
the questions that a vacuum selection principle, if it exists, would need
to answer.

\noindent
{\it Note added:}  As we were preparing to submit this paper,
Conlon and Quevedo posted an article~\cite{Conlon:Quevedo}
which also discusses gauge group rank issues on the landscape.

\vskip .25in
\noindent
{\bf Acknowledgments:}
We are grateful to K. Dienes, M. Dine, M. Douglas, G. Kane,
B. McNees, G. Shiu, E. Silverstein
and S. Thomas for helpful
discussions regarding the landscape.  We are especially
grateful to S. Kachru
for essential discussions.  This work is supported by the
Michigan Center for Theoretical Physics and the
Department of Energy.


\begin{thebibliography}{9}

\bibitem{flux1}
A.~Strominger,
``Superstrings With Torsion,''
Nucl.\ Phys.\ B {\bf 274}, 253 (1986).
J.~Polchinski and A.~Strominger,
``New Vacua for Type II String Theory,''
Phys.\ Lett.\ B {\bf 388}, 736 (1996)
[hep-th/9510227].
K.~Becker and M.~Becker,
``M-Theory on Eight-Manifolds,''
Nucl.\ Phys.\ B {\bf 477}, 155 (1996)
[hep-th/9605053].
J.~Michelson,
``Compactifications of type IIB strings to four dimensions with  non-trivial
classical potential,''
Nucl.\ Phys.\ B {\bf 495}, 127 (1997)
[hep-th/9610151].


\bibitem{flux2}
T.~R.~Taylor and C.~Vafa,
``RR flux on Calabi-Yau and partial supersymmetry breaking,''
Phys.\ Lett.\ B {\bf 474}, 130 (2000)
[hep-th/9912152].
B.~R.~Greene, K.~Schalm and G.~Shiu,
``Warped compactifications in M and F theory,''
Nucl.\ Phys.\ B {\bf 584}, 480 (2000)
[hep-th/0004103].
G.~Curio, A.~Klemm, D.~Lust and S.~Theisen,
``On the vacuum structure of type II string compactifications on  Calabi-Yau
spaces with H-fluxes,''
Nucl.\ Phys.\ B {\bf 609}, 3 (2001)
[hep-th/0012213].

\bibitem{Dasgupta:1999ss}
K.~Dasgupta, G.~Rajesh and S.~Sethi,
``M theory, orientifolds and G-flux,''
JHEP {\bf 9908}, 023 (1999)
[hep-th/9908088].

\bibitem{Giddings:2001yu}
S.~B.~Giddings, S.~Kachru and J.~Polchinski,
``Hierarchies from fluxes in string compactifications,''
Phys.\ Rev.\ D {\bf 66}, 106006 (2002)
[hep-th/0105097].

\bibitem{Kachru:2003aw}
S.~Kachru, R.~Kallosh, A.~Linde and S.~P.~Trivedi,
``De Sitter vacua in string theory,''
Phys.\ Rev.\ D {\bf 68}, 046005 (2003)
[hep-th/0301240].

\bibitem{Denef:2004dm}
F.~Denef, M.~R.~Douglas and B.~Florea,
``Building a better racetrack,''
JHEP {\bf 0406}, 034 (2004)
[hep-th/0404257].


\bibitem{Susskind:2003kw}
L.~Susskind,
``The anthropic landscape of string theory,''
hep-th/0302219.

\bibitem{Robbins:2004hx}
D.~Robbins and S.~Sethi,
``A barren landscape,''
arXiv:hep-th/0405011.


\bibitem{Douglas:2004qg}
M.~R.~Douglas,
``Statistical analysis of the supersymmetry breaking scale,''
hep-th/0405279.

\bibitem{Susskind:2004uv}
L.~Susskind,
``Supersymmetry breaking in the anthropic landscape,''
hep-th/0405189.


\bibitem{Dine:2004is}
M.~Dine, E.~Gorbatov and S.~Thomas,
``Low energy supersymmetry from the landscape,''
hep-th/0407043.

\bibitem{Silverstein:2004sh}
E.~Silverstein,
``Counter-intuition and scalar masses,''
hep-th/0407202.


\bibitem{Ashok:2003gk}
S.~Ashok and M.~R.~Douglas,
``Counting flux vacua,''
JHEP {\bf 0401}, 060 (2004)
[hep-th/0307049].

\bibitem{Denef:2004ze}
F.~Denef and M.~R.~Douglas,
``Distributions of flux vacua,''
JHEP {\bf 0405}, 072 (2004)
[hep-th/0404116].


\bibitem{Gukov:1999ya}
S.~Gukov, C.~Vafa and E.~Witten,
``CFT's from Calabi-Yau four-folds,''
Nucl.\ Phys.\ B {\bf 584}, 69 (2000)
[Erratum-ibid.\ B {\bf 608}, 477 (2001)]
[hep-th/9906070].

\bibitem{Giryavets:2004zr}
A.~Giryavets, S.~Kachru and P.~K.~Tripathy,
``On the taxonomy of flux vacua,''
JHEP {\bf 0408}, 002 (2004)
[hep-th/0404243].

\bibitem{Douglas:2003um}
M.~R.~Douglas,
``The statistics of string / M theory vacua,''
JHEP {\bf 0305}, 046 (2003)
hep-th/0303194].


\bibitem{Kachru:2002gs}
S.~Kachru, J.~Pearson and H.~Verlinde,
``Brane/flux annihilation and the string dual of a non-supersymmetric  field
theory,''
JHEP {\bf 0206}, 021 (2002)
[hep-th/0112197].

\bibitem{Gopakumar:1996mu}
R.~Gopakumar and S.~Mukhi,
``Orbifold and orientifold compactifications of F-theory and M-theory  to six
and four dimensions,''
Nucl.\ Phys.\ B {\bf 479}, 260 (1996)
[hep-th/9607057].



\bibitem{Marchesano:2004xz}
F.~Marchesano and G.~Shiu,
``Building MSSM Flux Vacua,''
hep-th/0409132.


\bibitem{Marchesano:2004yq}
F.~Marchesano and G.~Shiu,
``MSSM vacua from flux compactifications,''
hep-th/0408059.

\bibitem{Camara:2003ku}
P.~G.~Camara, L.~E.~Ibanez and A.~M.~Uranga,
``Flux-induced SUSY-breaking soft terms,''
Nucl.\ Phys.\ B {\bf 689}, 195 (2004)
[hep-th/0311241].

\bibitem{Krauss:1988zc}
L.~M.~Krauss and F.~Wilczek,
``Discrete Gauge Symmetry In Continuum Theories,''
Phys.\ Rev.\ Lett.\  {\bf 62}, 1221 (1989).
A.~Font, L.~E.~Ibanez and F.~Quevedo,
``Does Proton Stability Imply The Existence Of An Extra $Z^0$?,''
Phys.\ Lett.\ B {\bf 228}, 79 (1989).

\bibitem{Martin:1992mq}
S.~P.~Martin,
``Some simple criteria for gauged R-parity,''
Phys.\ Rev.\ D {\bf 46}, 2769 (1992)
[hep-ph/9207218].

\bibitem{Goh:2003nv}
R.~N.~Mohapatra,
``New Contributions To Neutrinoless Double-Beta Decay In Supersymmetric
Theories,''
Phys.\ Rev.\ D {\bf 34}, 3457 (1986);
H.~S.~Goh, R.~N.~Mohapatra, S.~Nasri and S.~P.~Ng,
``Proton decay in a minimal SUSY SO(10) model for neutrino mixings,''
Phys.\ Lett.\ B {\bf 587}, 105 (2004)
[hep-ph/0311330].

\bibitem{Aulakh:2000sn}
C.~S.~Aulakh, B.~Bajc, A.~Melfo, A.~Rasin and G.~Senjanovic,
``SO(10) theory of R-parity and neutrino mass,''
Nucl.\ Phys.\ B {\bf 597}, 89 (2001)
[hep-ph/0004031].

\bibitem{Dienes:1996du}
K.~R.~Dienes,
``String Theory and the Path to Unification: A Review of Recent Developments,''
Phys.\ Rept.\  {\bf 287}, 447 (1997)
[hep-th/9602045].


\bibitem{Ibanez:1991pr}
L.~E.~Ibanez and G.~G.~Ross,
``Discrete gauge symmetries and the origin of baryon and lepton number
conservation in supersymmetric versions of the standard model,''
Nucl.\ Phys.\ B {\bf 368}, 3 (1992).

\bibitem{Ibanez:1998qp}
L.~E.~Ibanez, R.~Rabadan and A.~M.~Uranga,
``Anomalous U(1)'s in type I and type IIB D = 4, N = 1 string vacua,''
Nucl.\ Phys.\ B {\bf 542}, 112 (1999)
[hep-th/9808139].

\bibitem{Ibanez:1999it}
L.~E.~Ibanez and F.~Quevedo,
``Anomalous U(1)'s and proton stability in brane models,''
JHEP {\bf 9910}, 001 (1999)
[hep-ph/9908305].

\bibitem{Banks:2003es}
T.~Banks, M.~Dine and E.~Gorbatov,
``Is there a string theory landscape?,''
JHEP {\bf 0408}, 058 (2004)
[hep-th/0309170].

\bibitem{Kachru:Aspen}
S. Kachru, ``Flux Compactifications'', talk delivered at
Aspen Center for Physics, 25 August 2004. O. DeWolfe,
A. Giryavets, S. Kachru and W. Taylor, to appear.

\bibitem{Sen:1997gv}
To understand the equivalences, see, e.g., A.~Sen,
``Orientifold limit of F-theory vacua,''
Phys.\ Rev.\ D {\bf 55}, 7345 (1997)
[hep-th/9702165],
and hep-th/9709159.

\bibitem{Conlon:Quevedo}
J.P.~Conlon, F.~Quevedo,
``On the explicit construction and statistics of
Calabi-Yau flux vacua,''
hep-th/0409215.












\end{thebibliography}
\end{document}